\documentclass[aps,prl,twocolumn,amsmath,amssymb,nofootinbib,showpacs,superscriptaddress]{revtex4}
\usepackage[english]{babel}
\usepackage{latexsym}
\usepackage{graphics}
\usepackage{subfigure}
\usepackage{epsfig}
\usepackage{color}

\begin{document}
\title{Crystalline phase of strongly interacting Fermi mixtures}

\author{D.S.~Petrov}
\affiliation{\mbox{Laboratoire de Physique Th\'{e}orique et Mod\`{e}les Statistiques, CNRS,
Universit\'{e} Paris Sud, 91405 Orsay, France}}
\affiliation{Russian Research Center Kurchatov Institute,
Kurchatov Square, 123182 Moscow, Russia}

\author{G.E.~Astrakharchik}
\affiliation{\mbox{Departament de F\'{\i}sica i Enginyeria Nuclear, Campus Nord B4-B5, Universitat Polit\`ecnica de Catalunya, E-08034 Barcelona, Spain}}

\author{D.J.~Papoular}
\affiliation{\mbox{Laboratoire de Physique Th\'{e}orique et Mod\`{e}les Statistiques, CNRS, 
Universit\'{e} Paris Sud, 91405 Orsay, France}}

\author{C.~Salomon}
\affiliation{\mbox{Laboratoire Kastler Brossel, CNRS, Ecole Normale Sup\'erieure, 24 rue Lhomond, 75231 Paris, France}}

\author{G.V.~Shlyapnikov}
\affiliation{\mbox{Laboratoire de Physique Th\'{e}orique et Mod\`{e}les Statistiques, CNRS,
Universit\'{e} Paris Sud, 91405 Orsay, France}}
\affiliation{\mbox{Van der Waals-Zeeman Institute, University of Amsterdam, Valckenierstraat 65/67, 1018 XE Amsterdam, The Netherlands}}

\date{\today}

\begin{abstract}
  
We show that the system of weakly bound molecules of heavy and light fermionic atoms is characterized by a long-range 
intermolecular repulsion and can undergo a gas-crystal quantum transition if the mass ratio exceeds a critical value.
For the critical mass ratio above 100 obtained in our calculations, this crystalline order can be observed as a 
superlattice in an optical lattice for heavy atoms with a small filling factor. We also find that this novel system
is sufficiently stable with respect to molecular relaxation into deep bound states and to the process of trimer formation.  
\end{abstract}
\maketitle

The use of Feshbach resonances for tuning the interaction in two-component 
ultracold Fermi gases of $^6$Li or $^{40}$K has led to remarkable developments, such as the observation of superfluid behavior 
in the strongly interacting regime through vortex formation \cite{zw1}, and Bose-Einstein condensation of 
weakly bound molecules of fermionic atoms on the positive side of the resonance (the atom-atom scattering 
length $a > 0$) \cite{bec}. Being highly excited, these extremely large diatomic molecules are remarkably 
stable with respect to collisional relaxation into 
deep bound states, which is a consequence of the Pauli exclusion principle for identical fermionic atoms \cite{PSS}. 

Currently, a new generation of experiments is being set up for studying mixtures of different 
fermionic atoms, with the idea of revealing the influence of the mass difference on superfluid properties and 
finding novel types of superfluid pairing. Weakly bound heteronuclear molecules on the positive side of the 
resonance are unique objects \cite{PSSJ,Hamburg}, which should manifest collisional stability and can pave a way to 
creating ultracold dipolar gases.


So far it was believed that dilute Fermi mixtures should be in the gas phase, like Fermi gases of atoms 
in two different internal states. In this paper we find that the system of molecules of heavy (mass $M$) and light 
(mass $m$) fermions can undergo a phase transition to a crystalline phase. This is due to   
a repulsive intermolecular potential originating from the exchange of light fermions and inversely proportional to $m$. As 
the kinetic energy of the molecules has a prefactor $1/M$, above a certain mass ratio $M/m$ the system can crystallize. 

We show that the interaction potential in a sufficiently dilute system of molecules is equal to the sum 
of their pair interactions and then analyze the case where the motion of heavy atoms is confined to two dimensions, whereas the light 
fermions can be either 2D or 3D \cite{comment1}. We calculate the zero-temperature gas-crystal 
transition line using Diffusion Monte Carlo (DMC) method and draw the phase diagram in terms 
of the mass ratio and density. This phase transition resembles the one for the flux lattice melting 
in superconductors, where the flux lines are mapped onto a system of bosons interacting  
via a 2D Yukawa potential \cite{Nelson}. In this case the Monte Carlo studies \cite{Ceperley,Blatter} 
identified the first order liquid-crystal transition at zero and finite temperatures. Aside from the difference 
in the interaction potentials, a distinguished feature of our system is related to its stability. The molecules
can undergo collisional relaxation into deep bound states, or 
form weakly bound trimers. We analyze resulting limitations on the lifetime of 
the system.

We first derive the Born-Oppenheimer interaction potential in the system of $N$ molecules. In this approach 
the state of light atoms adiabatically adjusts itself to the set of heavy-atom coordinates
$\{{\bf R}\}=\{{\bf R}_1,...,{\bf R}_N\}$ and one calculates the wavefunction and energy of light fermions in the 
field of fixed heavy atoms. Omitting the interaction between light (identical) fermions, it is sufficient to 
find $N$ lowest single-particle eigenstates, and the sum of their energies will give the 
interaction potential for the molecules. For the interaction between light and heavy atoms 
we use the Bethe-Peierls approach \cite{Bethe} assuming that the motion of light atoms is free 
everywhere except for their vanishing distances from heavy atoms.

The wavefunction of a single light atom then reads:
\begin{equation}   \label{psi} 
\Psi(\{{\bf R}\},{\bf r})=\sum_{i=1}^{N}C_{i}G_{\kappa}({\bf r}-{\bf R}_i),
\end{equation}
where ${\bf r}$ is its coordinate, and the Green function $G_{\kappa}$ satisfies the equation 
$(-\nabla_{\bf r}^2 +\kappa^2)G_{\kappa}({\bf r})=\delta({\bf r})$. The energy of the state (\ref{psi}) equals 
$\epsilon=-\hbar^2\kappa^2/2m$, and here we only search for negative single-particle energies (see below). The dependence of 
the coefficients $C_{i}$ and $\kappa$ on $\{{\bf R}\}$ is obtained using the Bethe-Peierls boundary condition:
\begin{equation}      \label{BP}
\Psi(\{{\bf R}\},{\bf r})\propto G_{\kappa_0}({\bf r}-{\bf R}_i);
\,\,\,\,\,{\bf r}\rightarrow {\bf R}_i.
\end{equation}
Up to a normalization constant, $G_{\kappa_0}$ is the wavefunction of a bound state of a single molecule with 
energy $\epsilon_0=-\hbar^2\kappa_0^2/2m$ and molecular size $\kappa_0^{-1}$. From Eqs.~(\ref{psi}) and (\ref{BP})  
one gets a set of $N$ equations: $\sum_jA_{ij}C_j=0$, where $A_{ij}=\lambda(\kappa)\delta_{ij}+G_{\kappa}(R_{ij})(1-\delta_{ij})$, 
$R_{ij}=|{\bf R}_i-{\bf R}_j|$, and $\lambda(\kappa)=\lim_{r\rightarrow 0}[G_{\kappa}(r)-G_{\kappa_0}(r)]$. 
The single-particle energy levels are determined by the equation
\begin{equation}     \label{det}
{\rm det}\left[ A_{ij}(\kappa,\{{\bf R}\})\right]=0.
\end{equation}

For $R_{ij}\rightarrow \infty$, Eq.~(\ref{det}) gives an N-fold degenerate 
ground state with $\kappa=\kappa_0$. At finite large $R_{ij}$,  
the levels split into a narrow band. Given a small parameter 
\begin{equation}     \label{ksi}
\xi=G_{\kappa_0}(\tilde R)/\kappa_0 |\lambda'_\kappa(\kappa_0)|\ll 1,
\end{equation} 
where $\tilde R$ is a characteristic distance at which heavy atoms can approach each other, the bandwidth is 
$\Delta\epsilon\approx4|\epsilon_0|\xi\ll |\epsilon_0|$. 
It is important for the adiabatic approximation that
all lowest $N$ eigenstates have negative energies and are separated from the continuum by a gap $\sim |\epsilon_0|$. 

We now calculate the single-particle energies up to second order in $\xi$. To this order we write
$\kappa(\lambda)\approx \kappa_0+\kappa'_\lambda\lambda+\kappa''_{\lambda\lambda}\lambda^2/2$ and turn from 
$A_{ij}(\kappa)$ to $A_{ij}(\lambda)$:
\begin{equation}       \label{Aij}
A_{ij}=\lambda\delta_{ij}+[G_{\kappa_0}(R_{ij})+\kappa'_\lambda\lambda 
\partial G_{\kappa_0}(R_{ij})/\partial\kappa](1-\delta_{ij}),
\end{equation}
where all derivatives are taken at $\lambda=0$. Using $A_{ij}$~(\ref{Aij}) in Eq.~(\ref{det}) 
gives a polynomial of degree $N$ in $\lambda$. Its roots $\lambda_i$ give the light-atom energy spectrum 
$\epsilon_i=-\hbar^2\kappa^2(\lambda_i)/2m$. The total energy, $E=\sum_{i=1}^{N}\epsilon_i$, is then given by 
\begin{equation}  \label{E}
E=-(\hbar^2/2m)\Big[ N\kappa_0^2+2\kappa_0\kappa'_\lambda\sum_{i=1}^{N}\lambda_i 
+(\kappa\kappa'_\lambda)'_\lambda\sum_{i=1}^{N}\lambda_i^2\Big].
\end{equation}
Keeping only the terms up to second order in $\xi$ and using 
basic properties of determinants and polynomial roots we find that the first order terms vanish, and 
the energy reads $E=N\epsilon_0+(1/2)\sum_{i\neq j}U(R_{ij})$, where 
\begin{equation}    \label{U}
U(R)=-\frac{\hbar^2}{m}\Big[\kappa_0(\kappa'_\lambda)^2
\frac{\partial G_{\kappa_0}^2(R)}{\partial\kappa}+(\kappa\kappa'_\lambda)'_\lambda G_{\kappa_0}^2(R)\Big].
\end{equation}
Thus, up to second order in $\xi$ the interaction in the system of N molecules is the sum of binary 
potentials (\ref{U}). 

If the motion of light atoms is 3D, the Green function is 
$G_{\kappa}(R)=(1/4\pi R)\exp(-\kappa R)$, and $\lambda(\kappa)=(\kappa_0-\kappa)/4\pi$,
with the molecular size $\kappa_0^{-1}$ equal to the 3D scattering length $a$.
Equation (\ref{U}) then gives a repulsive potential
\begin{equation}     \label{Yukawa}
U_{3D}(R)=4|\epsilon_0|( 1-(2\kappa_0R)^{-1})\exp(-2\kappa_0R)/\kappa_0R,
\end{equation}
and the criterion (\ref{ksi}) reads $(1/\kappa_0R)\exp(-\kappa_0R)\ll 1$. 
For the 2D motion of light atoms we have $G_{\kappa}(R)=(1/2\pi)K_0(\kappa R)$ and 
$\lambda(\kappa)=-(1/2\pi)\ln(\kappa/\kappa_0)$, where $K_0$ is the decaying Bessel function,
and $\kappa_0^{-1}$ follows from \cite{comment1}. 
This leads to a repulsive intermolecular potential:
\begin{equation}     \label{U2D}
U_{2D}(R)=4|\epsilon_0|[\kappa_0RK_0(\kappa_0R)K_1(\kappa_0R)-K_0^2(\kappa_0R)],
\end{equation}
with the validity criterion $K_0(\kappa_0R)\ll 1$. In both cases, which we denote $2\times 3$ and $2\times 2$ 
for brevity, the validity criteria are well satisfied already for $\kappa_0 R\approx 2$. 
 
The Hamiltonian of the many-body system reads:
\begin{equation}     \label{Hmb}   
H=-(\hbar^2/2M)\sum_i\Delta_{{\bf R}_i}+(1/2)\sum_{i\neq
j}U(R_{ij}), 
\end{equation} 
and the state of the system is determined by two parameters: the mass ratio $M/m$ and
the rescaled 2D density $n\kappa_0^{-2}$. At a large $M/m$, the potential repulsion dominates over
the kinetic energy and one expects a crystalline ground state. For separations $R_{ij}<\kappa_0^{-1}$ the adiabatic approximation breaks down. 
However, the interaction potential $U(R)$ is strongly repulsive at larger distances. 
Hence, even for an average separation between heavy atoms, ${\bar R}$, close to $2/\kappa_0$, 
they approach each other at distances smaller than $\kappa_0^{-1}$ with a small tunneling probability  
$P\propto\exp (-\beta \sqrt{M/m})\ll 1$, where $\beta\sim 1$. We extended $U(R)$ to 
$R\lesssim \kappa_0^{-1}$ in a way providing a proper 
molecule-molecule scattering phase shift in vacuum and checked that the phase diagram 
for the many-body system is not sensitive to the choice of this extension. 

Using the DMC method \cite{Boronat} we solved the many-body problem at zero temperature. For each phase,
gaseous and solid, the state with a minimum energy was obtained in a statistically exact way. 
The lowest of the two energies corresponds to the ground state, the other phase being metastable.  
The phase diagram is displayed in Fig.~1. The guiding wave 
function was taken in the Nosanow-Jastrow form \cite{2Ddipoles}. 
Simulations were performed
with $30$ particles and showed that the solid phase is a 2D triangular lattice. 
For the largest density we checked that using more particles has little effect on the results.

\begin{figure}[h]
\label{PhaseDiagram}
\includegraphics[width=0.9\hsize]{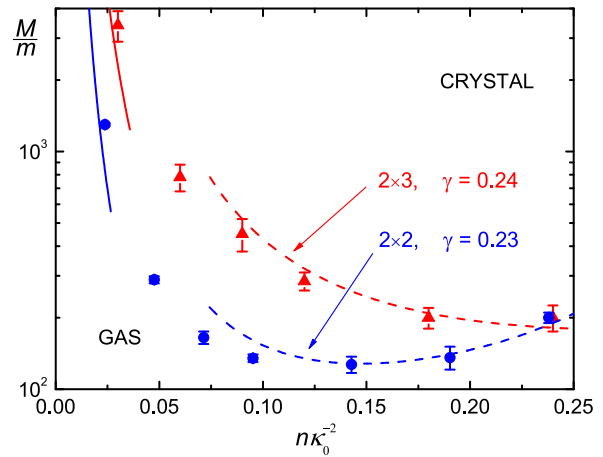}
\caption{
DMC gas-crystal transition lines for 3D (triangles) and 2D (circles) 
motion of light atoms. Solid curves show the low-density hard-disk limit, 
and dashed curves the results of the harmonic approach (see text).
}
\end{figure}

For both $2\times 3$ and $2\times 2$ cases the (Lindemann) ratio $\gamma$ of the rms displacement of molecules to  
${\bar R}$ on the transition lines ranges from $0.23$ to $0.27$.  
At low densities $n$ the de Broglie wavelength of molecules is 
$\Lambda\sim\gamma {\bar R}\gg\kappa_0^{-1}$, and $U(R)$ can be approximated by a hard-disk potential with the diameter
equal to the 2D scattering length. 
Then, using the DMC results for hard-disk bosons \cite{Xing}, we obtain the transition 
lines shown by solid curves in Fig.~1. At larger $n$, we have 
$\Lambda<\kappa_0^{-1}$ and use the
harmonic expansion of $U(R)$ around equlibrium positions in the crystal,
calculate the Lindemann ratio, and select $\gamma$ for the best
fit to the DMC data points (dashed curves in Fig.~1).

The mass ratio above 100, required for the observation of the crystalline order (see Fig.~1), can be achieved 
in an optical lattice with a small filling factor for heavy atoms. Their effective mass in the lattice, $M_*$,
can be made very large, and the discussed solid phase  
should appear as a superlattice. 
There is no interplay between the superlattice order and the shape of the underlying optical lattice,
in contrast to the recently studied solid and supersolid phases in a triangular lattice with the filling 
factor of order one \cite{TroyerSupersolid}.
Our superlattice remains compressible and supports two branches of phonons. 

The gaseous and solid phases of weakly bound molecules are actually
metastable. The main decay channels are the relaxation of molecules into deep bound states 
and the formation of trimer states by one light and two heavy atoms.  
A detailed analysis of scattering properties of these molecules will be given elsewhere, and here we
focus on their stability in an optical lattice. 

For a large effective mass ratio $M_*/m$, the relaxation into deep states occurs 
when a molecule is approached by another light atom and both light-heavy separations are 
of the order of the size of a deep state, $R_e\ll \kappa_0^{-1}$ \cite{comment2}. The released binding energy 
is taken by outgoing particles which escape from the sample.
The rate of this process is not influenced by the optical lattice.
 
We estimate this rate in the solid phase and near the gas-solid transition
to the leading order in $(\kappa_0{\bar R})^{-1}$.  
At light-heavy separations $r_{1,2}\ll\kappa_0^{-1}$ 
the initial-state wavefunction reads 
$\tilde\Psi=B(\kappa_0^{-1},{\bar R})\psi({\bf r}_1,{\bf r}_2)$.  
Writing it as an antisymmetrized product of wavefunctions (\ref{psi}),   
for the $2\times 3$ case ($\kappa_0^{-1}=a$) we find $B\approx(1/{\bar R}a^2)\exp(-{\bar R}/a)$. 
The quantity $W=B^2R_e^6$ is the probability of having both light atoms at distances $\sim R_e$
from a heavy atom, and the relaxation rate is $\nu_{3D}\propto W$. As the short-range physics is characterized by 
the energy scale $\hbar^2/mR_e^2$, we restore the dimensions and write
\begin{equation}     \label{nu3D}
\nu_{3D}=C(\hbar/m)(R_e/a)^4(1/{\bar R}^2)\exp(-2{\bar R}/a),
\end{equation}
where ${\bar R}^{-2}\approx n$. 
The coefficient $C$ depends on a particular system and is $\sim 1$ within an order of magnitude. 
The relaxation rate $\nu_{3D}$ is generally rather low. For K-Li mixture where $R_e\approx 50$\AA,
even at $na^2=0.24$ (see Fig.~1) the relaxation time exceeds 10 s for $n=10^9$ cm$^{-2}$ and $a=1600$\AA.
In the $2\times 2$ case, for the same $n$ and $\kappa_0^{-1}$ 
the probability $W$ is smaller and the relaxation is slower.

The formation of trimer bound states by one light and two heavy atoms occurs when two molecules approach each other
at distances $R\lesssim \kappa_0^{-1}$. It is accompanied by a release of the second light atom.
The existence of the trimer states is seen considering
a light atom interacting with two heavy ones. The lowest energy solution of Eq.~(\ref{det}) for $N=2$ is the 
gerade state ($C_1=C_2$). Its energy $\epsilon_{+}(R)$ introduces an effective attractive  
potential acting on the heavy atoms, and the trimer states are bound states of two heavy atoms in this potential.

In an optical lattice the trimers are eigenstates of the Hamiltonian 
$H_0=-(\hbar^2/2M_*)\sum_{i=1,2}\Delta_{\bf R_i}+\epsilon_{+}(R_{12})$. In a deep lattice one can 
neglect all higher bands and regard ${\bf R_i}$ as discrete lattice coordinates and $\Delta$ as the lattice Laplacian. 
Then, the fermionic nature of the heavy atoms prohibits them to be in the same lattice site. For a very 
large mass ratio $M_*/m$ the kinetic energy term in $H_0$ can be neglected, and the lowest trimer state has energy 
$\epsilon_{\rm tr}\approx \epsilon_{+}(L)$, where $L$ is the lattice period. It consists of a pair of heavy atoms 
localized at neighboring sites and a light atom in the gerade state. Higher trimer states are formed by 
heavy atoms localized in sites separated 
by distances $R>L$. This picture breaks down at large $R$, where the spacing between trimer levels
is comparable with the tunneling energy $\hbar^2/M_*L^2$ and the heavy atoms are delocalized. 

In the many-body molecular system the scale of energies in Eq.~(\ref{Hmb}) is much smaller than $|\epsilon_0|$. Thus, 
the formation of trimers in molecule-molecule ``collisions'' is energetically allowed only if the trimer binding energy 
is $\epsilon_{\rm tr}<2\epsilon_0$. Since the lowest trimer energy in the optical lattice is $\epsilon_{+}(L)$, the trimer formation
requires the condition $\epsilon_{+}(L)\lesssim 2\epsilon_0$, which is equivalent to $\kappa_0^{-1}\gtrsim 1.6 L$ in the 
$2\times 3$ case and $\kappa_0^{-1}\gtrsim 1.25 L$ in the $2\times 2$ case. This means that for a sufficiently small 
molecular size or large lattice period $L$ the formation of trimers is forbidden. 

At a larger molecular size or smaller $L$ the trimer formation is possible. For finding the rate we consider the
interaction between two molecules as a reduced 3-body problem, accounting for the fact that one of the light atoms is in the gerade
and the other one in the ungerade state ($C_1=-C_2$). The gerade light atom is integrated out and is 
substituted by the effective potential $\epsilon_+(R)$. For the ungerade state the adiabaticity breaks down at 
inter-heavy separations $R\lesssim \kappa_0^{-1}$, and the ungerade light atom is treated explicitly. The wavefunction of the reduced 3-body 
problem satisfies the Schr\"odinger equation
\begin{equation}\label{Schr}
[H_0-\hbar^2\nabla_{\bf r}^2/2m-E]\psi(\{{\bf R}\},{\bf r})=0,
\end{equation}
where the energy $E$ is close to $2\epsilon_0$, $\{{\bf R}\}$ denotes the set $\{{\bf R}_1,{\bf R}_2\}$, and ${\bf r}$ is the 
coordinate of the ungerade atom. 
The interaction between this atom and the heavy ones is replaced by the boundary condition (\ref{BP}) on $\psi$. The 3-body 
problem can then be solved by encoding the information on 
the wavefunction $\psi$ in an auxiliary function $f(\{\tilde{\bf R}\})$ \cite{Petrov3BodyFermions} and representing 
the solution of Eq.~(\ref{Schr}) in the form:
\begin{equation}\label{reducedpsi}
\psi=\sum_{\{\tilde{\bf R}\},\nu}\chi_{\nu}(\{{\bf R}\})\chi_{\nu}^{*}(\{\tilde{\bf R}\})
f(\{\tilde{\bf R}\})F_{\kappa_\nu}({\bf r},\{\tilde{\bf R}\}),
\end{equation}
where $\chi_\nu (\{{\bf R}\})$ is an eigenfunction of $H_0$ with energy $\epsilon_\nu$, and 
$F_{\kappa_\nu}({\bf r},\{\tilde{\bf R}\})=G_{\kappa_\nu}({\bf r}-\tilde{\bf R}_1)-G_{\kappa_\nu}({\bf r}-\tilde{\bf R}_2)$ with 
$\kappa_\nu=\sqrt{2m(\epsilon_\nu-E)/\hbar^2}$.   
For $\epsilon_\nu<E$ the trimer formation in the state $\nu$ is possible. This is consistent with 
imaginary $\kappa_\nu$ and the Green function $G_{\kappa_\nu}$ describing an outgoing wave of the light atom and trimer. 

We derive an equation for the function $f$ in a deep lattice, where the tunneling energy $\hbar^2/M_*L^2\ll |\epsilon_0|$.
Then the main contribution to the sum in Eq.~(\ref{reducedpsi}) comes from the states 
$\nu$ for which $|\epsilon_{\nu}-\epsilon_+(R_{12})|\lesssim\hbar^2/M_*L^2$. The sum is calculated by
expanding $\kappa_{\nu}$ around $\kappa(R_{12})=\sqrt{2m(\epsilon_+(R_{12})-E)/\hbar^2}$ up 
to first order in $(\epsilon_{\nu}-\epsilon_+(R_{12}))/\epsilon_0$
and using the equation $(H_0-\epsilon_{\nu})\chi_{\nu}=0$. 
The equation for $f$ is then obtained by taking the limit ${\bf r}\rightarrow {\bf R}_1$ in the resulting expression for 
$\psi$ and comparing it with the boundary condition (\ref{BP}).
This yields
\begin{equation}\label{FinalSchr} 
[-(\hbar^2/2M_*)\sum_{i=1,2}\Delta_{\bf R_i} + U_{\rm eff}(R_{12})]f({\bf R}_1,{\bf R}_2)=0,
\end{equation} 
where the effective potential $U_{\rm eff}(R_{12})$ is given by
\begin{equation}\label{FinalUeff}
U_{\rm eff}(R)=\frac{\hbar^2\kappa(R)}{m}\frac{\lambda(\kappa(R))-G_{\kappa(R)}(R)}{(\partial/\partial \kappa)[\lambda(\kappa(R))-G_{\kappa(R)}(R)]}.
\end{equation}

At large distances one has $U_{\rm eff}\approx U(R)+2\epsilon_0-E$, and for smaller $R$ where $\epsilon_{+}(R)<E$, 
the potential $U_{\rm eff}$ acquires an imaginary part accounting for the decay of molecules into trimers.  
The number of trimer states that can be formed grows with the 
molecular size. Eventually it becomes independent of $L$ and so does the loss rate.

In this limit, we solve Eq.~(\ref{FinalSchr}) for two molecules with zero total momentum 
under the condition that $f({\bf R}_1,{\bf R}_2)$ is maximal for $|{\bf R}_1-{\bf R}_2|={\bar R}\approx n^{-1/2}$. 
We thus obtain $E$ as a function of the density and mass ratio, and its imaginary part gives the loss rate $\nu$ for the many body system. Numerical analysis for 
$0.06<n\kappa_0^{-2}<0.4$ and $50<M_*/m<2000$ is well fitted by $\nu\approx (D\hbar n/M_*)(n\kappa_0^{-2})\exp(-J\sqrt{M_*/m})$, with $D=7$ and
$J=0.95-1.4(n\kappa_0^{-2})$ for the $2\times 3$ case, and $D=10^2$, $J=1.45-2.8(n\kappa_0^{-2})$ in the $2\times 2$ case. 
One can suppress $\nu$ by increasing $M_*/m$, whereas for $M_*/m\lesssim 100$ the trimers can be formed on a time scale 
$\tau\lesssim 1$ s. 

In conclusion, we have shown that the system of weakly bound molecules of heavy and light fermionic atoms can undergo a 
gas-crystal quantum transition. The necessary mass ratio is above 100 and the observation of such crystalline order requires 
an optical lattice for heavy atoms, where it appears as a superlattice. 
A promising candidate is the $^6$Li-$^{40}$K mixture as the Li atom may tunnel freely in a lattice while localizing 
the heavy K atoms to reach high mass ratios. A lattice with 
period 250 nm and K effective mass M*= 20M provide a tunneling 
rate $\sim 10^3$ s$^{-1}$ sufficiently fast to let the crystal form. Near 
a Feshbach resonance, a value $a=500$ nm gives 
a binding energy $300$ nK, and lower temperatures  
should be reached in the gas. The parameters $n\kappa_0^{-2}$ of Fig.~1 are then 
obtained at 2D densities in the range $10^7-10^8$ cm$^{-2}$ easily reachable in 
experiments. For $n=10^8$ cm$^{-2}$ the rate of the trimer formation is of the order of seconds, and these 
peculiar bound states can be detected optically.

This work was supported by the IFRAF Institute, by
ANR (grants 05-BLAN-0205, 05-NANO-008-02, and 06-NANO-014), by the Dutch Foundation FOM, by the Russian Foundation for Fundamental
Research, and by the National Science Foundation (Grant No. PHY05-51164). 
LKB is a research unit no.8552 of CNRS, ENS, and University
of Pierre et
Marie Curie. LPTMS is a research unit no.8626 of CNRS and University
Paris-Sud. 



\begin{thebibliography}{99}

\bibitem{zw1} M.W. Zwierlein {\it et al.}, Nature (London) {\bf 435}, 1047 (2005).

\bibitem{bec} M. Greiner {\it et al.}, Nature (London) {\bf 426}, 537 (2003); S. Jochim {\it et al.}, 
Science {\bf 302}, 2101 (2003); M. W. Zwierlein {\it et al.}, Phys. Rev. Lett. {\bf 91}, 250401 (2003); 
T. Bourdel {\it et al.}, {\it ibid.} {\bf 93}, 050401 (2004); G.B. Partridge, {\it et al.}, 
{\it ibid.} {\bf 95}, 020404 (2005).

\bibitem{PSS} D. S. Petrov, C. Salomon, and G. V. Shlyapnikov, Phys. Rev. Lett. {\bf 93}, 090404 (2004); Phys. Rev. A {\bf 71}, 012708 (2005).

\bibitem{PSSJ} D.S. Petrov, C. Salomon, and G.V. Shlyapnikov, J. Phys. B 
{\bf 38}, S645 (2005). 

\bibitem{Hamburg} C. Ospelkaus {\it et al.}, Phys. Rev. Lett. {\bf 97}, 120402 (2006).

\bibitem{comment1} In the 2D regime achieved by confining the light-atom motion to zero
point oscillations with amplitude $l_0$, the weakly bound molecular states exist at a 
negative $a$ satisfying the condition $|a|\ll l_0$. See 
D.S. Petrov and G.V. Shlyapnikov, Phys. Rev. A {\bf 64}, 012706 (2001).

\bibitem{Nelson} D.R. Nelson and H.S. Seung, Phys. Rev. B {\bf 39}, 9153 (1989).

\bibitem{Ceperley} W.R. Magro and D.M. Ceperley, Phys. Rev. B {\bf 48}, 411 (1993).

\bibitem{Blatter} H. Nordborg and G. Blatter, Phys. Rev. Lett. {\bf 79}, 1925 (1997).

\bibitem{Bethe} H. Bethe and R. Peierls, Proc. R. Soc. London, Ser. A
{\bf 148}, 146 (1935).

\bibitem{Boronat} For a general reference on the DMC method see, {\it e.g.},
J. Boronat and J. Casulleras, Phys. Rev. B {\bf 49}, 8920 (1994).

\bibitem{2Ddipoles} G.E. Astrakharchik {\it et al.},  
Phys. Rev. Lett. {\bf 98}, 060405 (2007)

\bibitem{Xing} L. Xing, Phys. Rev. B {\bf 42}, 8426 (1990).

\bibitem{TroyerSupersolid} S. Wessel and M. Troyer, Phys. Rev. Lett. {\bf 95}, 127205 (2005); 
D. Heidarian and K. Damle, {\it ibid.} {\bf 95}, 127206 (2005); 
R.G. Melko {\it et al.}, {\it ibid.} {\bf 95}, 127207 (2005).


\bibitem{comment2} The relaxation involving one light and two heavy atoms
is strongly suppressed as it requires the heavy atoms to approach each other and get to the same lattice site. 

\bibitem{Petrov3BodyFermions} D.S. Petrov, Phys. Rev. A {\bf 67}, 010703(R) (2003).





\end{thebibliography}
\end{document}